\documentclass[12pt]{article}
\usepackage{graphics}
\usepackage{amsmath}
\usepackage{amssymb}

\makeatletter\@addtoreset{equation}{section}
\makeatother
\begin{document}
\begin{titlepage}
\begin{flushright}
TIT/HEP-584\\
May, 2008\\
\end{flushright}
\vspace{0.5cm}
\begin{center}
{\Large \bf 
A Numerical Study of Gluon Scattering Amplitudes in ${\cal N}=4$ 
Super Yang-Mills Theory at Strong Coupling
}
\lineskip .75em
\vskip1.0cm
{\large Suguru Dobashi${}^{1}$, Katsushi Ito${}^{2}$ and 
Koh Iwasaki${}^{2}$ }
\vskip 2.5em
${}^{1}$ {\normalsize\it 
Department of Accelerator and Medical Physics,\\
National Institute of Radiological Sciences, \\
4-9-1 Anagawa, Inage-ku, Chiba 263-8555, Japan
} \vskip 1.5em
${}^{2}$ {\normalsize\it Department of Physics\\
Tokyo Institute of Technology\\
Tokyo, 152-8551, Japan}  

\vskip 4.5em
\end{center}
\begin{abstract}
We study gluon scattering amplitudes in ${\cal N}=4$ super Yang-Mills
 theory at strong coupling via the AdS/CFT correspondence.
We solve numerically 
the discretized Euler-Lagrange equations on the square worldsheet 
for the minimal surface with light-like boundaries in AdS spacetime.
We evaluate the area of the surface for the 4, 6 and 8-point amplitudes
using worldsheet and radial cut-off regularizations.
Their infrared singularities in the cut-off regularization 
are found to agree with the analytical results
near the cusp less than 5\% at 520$\times$520 lattice points.

\end{abstract}
\end{titlepage}

\baselineskip=0.7cm
\section{Introduction}
Recently Alday and Maldacena \cite{AlMa} proposed the prescription 
for the calculation of gluon scattering amplitudes in 
${\cal N}=4$ super Yang-Mills (SYM) theory
at strong coupling via the AdS/CFT correspondence.
They showed that the gluon scattering amplitude is related
to the area of a minimal surface in AdS spacetime surrounded by the 
Wilson loop with light-like boundaries.
They found the exact solution of the minimal surface 
corresponding to the 4-point amplitude and showed that 
it reproduces the
perturbative formula conjectured by Bern, Dixon and Smirnov (BDS)
\cite{BeDiSm}.

The correspondence between gluon amplitudes and the Wilson loops has been
examined also at weak coupling\cite{Wilson}. 
It is shown in \cite{DrHeKoSo} that the Wilson loops at weak coupling 
obey the anomalous conformal Ward identity, which determines the 
$n=4$ and $5$-point amplitudes completely.
For $n\geq 6$-point amplitudes, however, the conformal invariance of the 
amplitudes does not fix their dependence on the kinematical variables. 
Recently it is found that the 6-point 2-loop corrections to the Wilson loop
agrees  numerically with the gluon amplitudes but they 
differ from the BDS formula\cite{BeDiKoRoSpVeVo}.

In order to study gluon scattering amplitudes at strong coupling using
the AdS/CFT correspondence, 
we need to find the solution of the minimal surface in AdS spacetime
surrounded by the light-like Wilson loop.
See 
\cite{Bu,MiMoTo, AlMa2, AsDoItNa, KoRa, ItNaIw,OzThYa, 
gluons} for references.
The minimal surface is obtained by solving the Euler-Lagrange equations with 
the Dirichlet boundary conditions.
These are non-linear partial 
differential equations and highly non-trivial to solve analytically
for the polygon with $n\geq 5$ boundaries. 
Any exact solution for $n\geq 5$ is not yet known so far.
In a previous paper \cite{AsDoItNa}, the authors including the two 
authors in the present paper, constructed solutions for
the 6 and 8-point amplitudes
by cutting and gluing the 4-point amplitude.
The evaluation of the amplitudes shows that they do not agree with the 
BDS formula both for the infrared singularity and finite parts.
In particular, the infrared singularities coincide with those of the
4-point amplitudes, which suggests that these simple solutions
 are not the minimal surfaces and 
might correspond to the other disconnected amplitudes. 

In this paper we will propose a practical approach to  
compute the minimal surface in AdS spacetime.
We investigate  numerically the solution of
the Euler-Lagrange
 equations for the minimal surface in AdS spacetime, surrounded
by the light-like segments.
We solve the discretized Euler-Lagrange equations on the square 
lattice with the
Dirichlet boundary conditions.
We evaluate the area of the surface for the 4, 6 and 8-point amplitudes
using two types of cut-off regularizations.
One is the world-sheet cut-off regularization. The other is the 
radial cut-off regularization \cite{Al}.
For the 4-point amplitude we see that the numerical result agrees with the
analytical solution obtained by Alday and Maldacena.
For the 6 and 8-point solutions, we take the same momenta configuration 
as in \cite{AsDoItNa}. 
The results are different from the cut and glue solutions and there appear
new cusp singularities.
Their infrared singularities are found to agree with analytical
solutions near the cusp.
This numerical approach would be a useful method to calculate the $n$-point
gluon amplitudes at strong coupling and test the BDS conjecture from the 
AdS side.

The paper is organized as follows.
In Section 2 we describe the discretized Euler-Lagrange equations in 
AdS spacetime and solve the equations numerically.
In Section 3, we evaluate the area of the minimal surface for 4, 6 
and 8-point
 amplitudes by using 
 the cut-off regularization in the radial direction of
AdS spacetime \cite{Al}. We compare their infrared
singularity part with the analytical results near the cusp.
Section 5 includes conclusions and some comments.

\section{The Euler-Lagrange equations of the minimal surface}
\subsection{Analytical solutions}
In this paper we investigate the minimal surface in AdS$_5$ 
surrounded by the curve $C_n$ made of 
light-like segments $\Delta y^\mu=2\pi p_i^\mu$, 
which corresponds to the $n$-point gluon amplitude
with on-shell momenta $p_i$ ($p^2_i=0$, $i=1,\cdots, n$).
Here $y^\mu$ ($\mu=0,1,2,3$) together with the radial coordinate
 $r$ are the Poincar\'e coordinates 
in AdS$_5$ spacetime with the metric
\begin{equation}
ds^2=R^2{dy^\mu dy_\mu+dr^2\over r^2},
\end{equation}
and $R$ is the radius of AdS$_5$.
It is convenient to write the Nambu-Goto action in the static gauge 
where we put $y_3=0$ and parametrize the surface by 
$y_1$ and $y_2$.
Then $r$ and $y_0$ are functions of $y_1$ and $y_2$ and the action is
given by
\begin{equation}
 S={R^2\over 2\pi}
\int dy_1 dy_2 \frac{\sqrt{1+(\partial_i r)^2-(\partial_i y_0)^2
-(\partial_1 r \partial_2 y_0-\partial_2 r \partial_1 y_0)^2}}{r^2}.
\label{eq:ngaction}
\end{equation}
The Euler-Lagrange equations become
\begin{eqnarray}
&&  \partial_i
\left(
\frac{\partial L}{\partial(\partial_i y_0)}
\right)=0, 
\quad \partial_i
\left(
\frac{\partial L}{\partial (\partial_i r)}
\right)-\frac{\partial L}{\partial r}=0,
\label{eq:euler1}
\end{eqnarray}
where $L$ is the Lagrangian of the action.
The minimal surface is obtained by solving the Euler-Lagrange equations 
(\ref{eq:euler1}), which  are non-linear partial equations
and difficult to solve analytically.

For $n=4$, Alday and Maldacena found the exact
solution \cite{AlMa}.
Let us consider the scattering amplitude for two incoming particles 
with momenta $p_1$
and $p_3$ and outgoing particles with momenta $p_2$ and $p_4$.
The Mandelstam variables are given by $s=-(p_1+p_2)^2$
and $t=-(p_2+p_3)^2$.
The case with $s=t$ is particularly simple where the Wilson loop is 
represented by the square with corners at $y_1,y_2=\pm 1$. 
The momenta in the $(y_0,y_1,y_2)$-space associated with the square are
\begin{eqnarray}
&& 2\pi p_1=(2,2,0), \quad 2\pi p_2=(-2,0,2), \quad 2\pi p_3=(2,-2,0),
  \quad
2\pi p_4=(-2,0,-2).\nonumber\\
\label{eq:4ptmom1}
\end{eqnarray}
The boundary conditions for the case with $s=t$ are given by
\begin{equation}
r(\pm 1, y_2)=r(y_1,\pm 1)=0,\quad
y_{0}(\pm 1, y_2)=\pm y_2, \quad
y_{0}(y_1,\pm 1)=\pm y_1.
\label{eq:4ptbc}
\end{equation}
The solution of the Euler-Lagrange
 equations satisfying the boundary conditions is 
\begin{equation}
 y_0(y_1,y_2)=y_1 y_2,\quad r(y_1,y_2)=\sqrt{(1-y_1^2)(1-y_2^2)}.
\label{eq:4pt}
\end{equation}
By applying the conformal transformation $SO(2,4)$ one can obtain
the solution with general $s$ and $t$.
Using the dimensional regularization,  the  area is shown to 
agree with the BDS formula at strong coupling.

Noticing that the solution (\ref{eq:4pt}) satisfies
eq.~(\ref{eq:euler1}) even when we change the sign of $y_0$,
an obvious and simple generalization of this remarkable 4-point solution is 
to cut and glue the solution.
In \cite{AsDoItNa}, we have constructed 6-point and 8-point solutions
of the same action.
These cut and glue solutions are  summarized as follows:
\paragraph{6-point function solution 1:}
\begin{equation}
 y_0={1\over2}(|y_1 y_2|+y_1 y_2-|y_1| y_2 +y_1 |y_2|),\quad
r=\sqrt{(1-y_1^2)(1-y_2^2)}.
\label{eq:6pt1}
\end{equation}
The solution corresponds to the scattering with momenta 
\begin{eqnarray}
 2\pi p_1&=&(2,0,-2),\quad
 2\pi p_2=(-1,1,0),\quad
2\pi p_3=(1,1,0),\nonumber\\
2\pi p_4&=&(-1,0,1),\quad
 2\pi p_5=(1,0,1),\quad
2\pi p_6=(-2,-2,0).
\end{eqnarray}

\paragraph{6-point function solution 2:}
\begin{equation}
 y_0=y_1 |y_2|, \quad
r=\sqrt{(1-y_1^2)(1-y_2^2)}.
\label{eq:6pt2}
\end{equation}
The momenta are
\begin{eqnarray}
 2\pi p_1&=&(1,0,-1),\quad
 2\pi p_2=(-1,0,-1),\quad
2\pi p_3=(2,2,0), \nonumber\\
2\pi p_4&=&(-1,0,1),\quad
 2\pi p_5=(1,0,1),\quad
2\pi p_6=(-2,-2,0).
\end{eqnarray}
\paragraph{8-point function:}
\begin{equation}
 y_0=| y_1 y_2|, \quad
r=\sqrt{(1-y_1^2)(1-y_2^2)}.
\label{eq:8pt}
\end{equation}
The momenta are
\begin{eqnarray}
 2\pi p_1&=&(-1,0,-1),\quad
 2\pi p_2=(1,0,-1),\quad
2\pi p_3=(-1,1,0), \nonumber\\
2\pi p_4&=&(1,1,0),\quad
 2\pi p_5=(-1,0,1),\quad
2\pi p_6=(1,0,1),\nonumber\\
2\pi p_7&=&(-1,-1,0),\quad
 2\pi p_8=(1,-1,0).
\end{eqnarray}
These solutions  do not reproduce the BDS formula.
The most significant discrepancy appears in their infrared
singularities. 
In fact, these solutions have the same infrared 
divergences as the 4-point amplitude.
These solutions have also the delta function source term in the 
equation of motion.
It has been discussed in \cite{AsDoItNa} 
that these are not the minimal surface and correspond to other
disconnected diagrams in field theory.

The main purpose of this paper is to explore numerically 
the minimal surface
corresponding to the same boundary conditions as the above higher-point 
solutions.
In this paper we will study the square worldsheet for simplicity.
We will make a comment on a generalization to the light-like hexagon 
Wilson loop in Sect. 4.

\subsection{Discretized Euler-Lagrange Equations}
We firstly introduce the square lattice with spacing $h=\frac{2}{M}$
where $M$ is a positive integer.
At each site $(-1+h i,-1+h j)$ ($i,j=0,\cdots, M$), we assign the variables
\begin{equation}
 y_0[i,j]=y_0(-1+h i,-1+h j),\quad
r[i,j]=r(-1+h i,-1+h j).
\end{equation}
We use the central difference method to discretize the Euler-Lagrange
 equations.
Namely, for a function $f(y_1,y_2)$ we adopt the following rules to
obtain the difference equations from the differential equations:
\begin{eqnarray}
 \partial_1 f(y_1,y_2)&\longrightarrow& \frac{f[i+1,j]-f[i-1,j]}{2h},
\nonumber\\
 \partial_2 f(y_1,y_2)&\longrightarrow& \frac{f[i,j+1]-f[i,j-1]}{2h},
\nonumber\\
\partial_1 \partial_2 f(y_1,y_2)&\longrightarrow& 
\frac{f[i+1,j+1]-f[i+1,j-1]-f[i-1,j+1]+f[i-1,j-1]}{4h^2},
\nonumber\\
\partial_1^2 f(y_1,y_2)&\longrightarrow& 
\frac{f[i+1,j]-2f[i,j]+f[i-1,j]}{h^2},
\nonumber\\
\partial_2^2 f(y_1,y_2)&\longrightarrow& 
\frac{f[i,j+1]-2f[i,j]+f[i,j-1]}{h^2}.
\end{eqnarray}
We need to specify the boundary conditions to solve the equations.
For example, the boundary conditions (\ref{eq:4ptbc}) for the 
4-point solutions lead to
\begin{eqnarray}
 y_0[i,0]&=& y_0(-1+hi,-1), \quad  y_0[i,M]=y_0(-1+hi,1),
\nonumber\\
 y_0[0,j]&=& y_0(-1,-1+hj), \quad  y_0[M,j]=y_0(1,-1+hj),
\nonumber\\
r[i,0]&=&r[i,M]=r[0,j]=r[M,j]=0.
\end{eqnarray}
Then we obtain $2\times (M-1)^2$ nonlinear simultaneous equations for
$y_0[i,j]$ and $r[i,j]$.
We will use Newton's method to find a numerical solution. 
In this method the initial solution is important to get good 
numerical results iteratively.
The approximate
solution will soon converge if the initial numerical data is
appropriate.
Otherwise, the numerical solution does not converge and the surface
would not be smooth.
We take as the initial condition 
the Alday-Maldacena solution for 4-point function 
or cut and glue solutions for 6 and 8-point functions.
But some trials show that the cut and glue solution 
does not lead to convergence
when the lattice size becomes large.
So we begin with the $M=10$ lattice with these initial conditions
and proceed step by step to larger size of lattices, up to 
$M=520$\footnote{A numerical computation was carried out by 
Mathematica and C
programs on a personal
computer with 8GB memory, which restricts the lattice size to $M=520$.},
with the output of a previous smaller lattice calculation 
being the input for the next larger lattice,
by linearly interpolating the output.
For each lattice size,
the Newton's method is repeatedly applied until
the non-linear simultaneous equation is satisfied 
up to ${\cal O}(10^{-16})$ and 
the area of the obtained surface
does not change up to ${\cal O}(10^{-6})$
even when we proceed to the next step of the Newton's method.
The area is approximately evaluated as $S=\sum L[i,j]h^2$, 
where $L[i,j]$ and $h$ are the discretized Lagrangian at 
a lattice point $(i,j)$ and the lattice spacing, respectively.

\begin{figure}[bthp]
\begin{center}
\begin{tabular}{cc}
\resizebox{75mm}{!}{\includegraphics{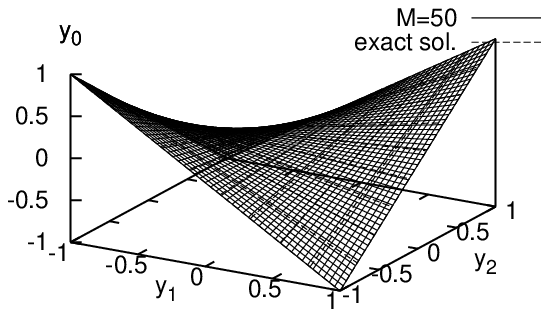}} &
\resizebox{75mm}{!}{\includegraphics{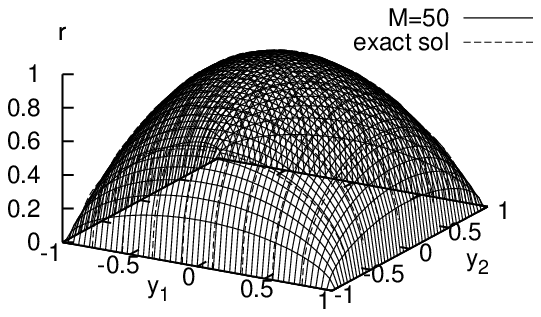}}\\
(a) & (b)
\end{tabular}
\end{center}
\caption{
Numerical solution of the 4-point amplitude at $M=50$.
}\label{fig:4pt}
\end{figure}

\begin{figure}[bthp]
\begin{center}
\begin{tabular}{cc}
\resizebox{75mm}{!}{\includegraphics{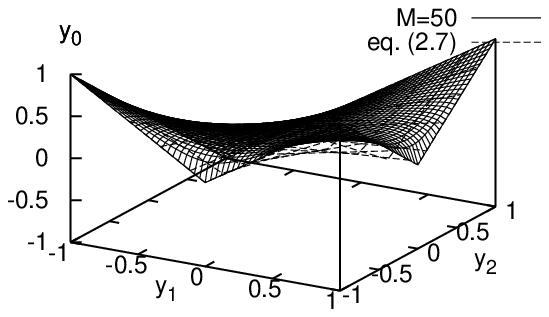}} &
\resizebox{75mm}{!}{\includegraphics{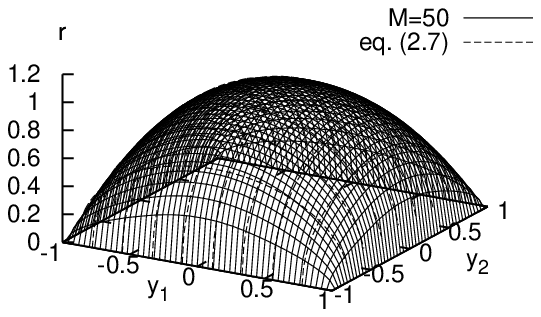}}
\\
(a) & (b)
\end{tabular}
\end{center}
\caption{
Numerical solution of 6-point  amplitude solution 1 (a) $y_0$ (b) $r$
at $M=50$.
}\label{fig:6pt1}
\end{figure}

\begin{figure}[bthp]
\begin{center}
\begin{tabular}{cc}
\resizebox{75mm}{!}{\includegraphics{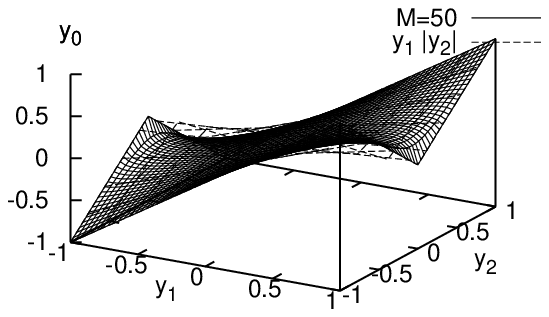}} &
\resizebox{75mm}{!}{\includegraphics{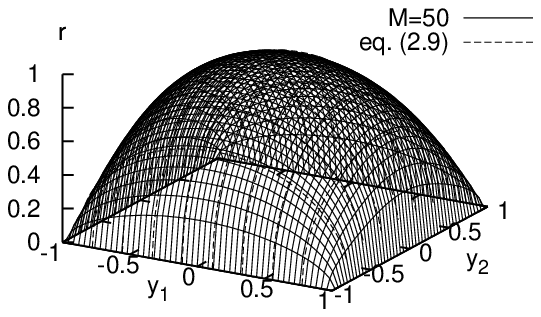}}
\\
(a) & (b)
\end{tabular}
\end{center}
\caption{
Numerical solution of 6-point  amplitude solution 2 (a) $y_0$ (b) $r$
at $M=50$.
}\label{fig:6pt2}
\end{figure}

\begin{figure}[bthp]
\begin{center}
\begin{tabular}{cc}
\resizebox{75mm}{!}{\includegraphics{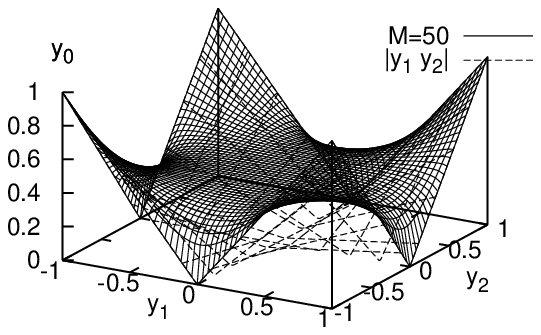}} &
\resizebox{75mm}{!}{\includegraphics{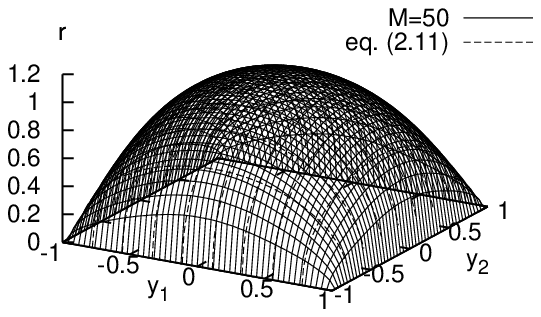}}
\\
(a) & (b)
\end{tabular}
\end{center}
\caption{
Numerical solution of 8-point amplitude (a) $y_0$ (b) $r$
at $M=50$.
}\label{fig:8pt}
\end{figure}

We plot numerical solutions at $M=50$ in Fig.~\ref{fig:4pt}, where we 
can see the 4-point solution apparently agrees with the exact solution.
This agreement with the exact solution is quantitatively checked
by comparing the areas between numerical and exact solutions
in the next section.
In Figs.~\ref{fig:6pt1}, \ref{fig:6pt2} and \ref{fig:8pt}, we plot numerical 
solutions of the 6-point and
8-point amplitudes with the same boundary conditions as the cut and glue
 solutions.
We find that 
the numerical solutions are different from the cut and glue
solutions and new cusps seem to appear which are absent in the cut and
glue solutions \cite{AsDoItNa}.
In the next section 
we will evaluate the 
area of the surface numerically and examine the infrared behavior of the
regularized area.

\section{The cut-off regularization and the minimal area}
In order to study the gluon scattering amplitude, we need to 
evaluate the area of the minimal surface in AdS spacetime.
Since we have obtained numerically the solution of the 
surface from the discretized Euler-Lagrange equations, 
we can evaluate it easily.
But for a light-like Wilson loop, the value of the area diverges due to 
the cusp \cite{Kr, AlMa}.
We often use the dimensional regularization scheme to control this
divergence,
which is convenient to compare the area with the field theory result.
But in order to calculate the area numerically in the present
discretization procedure, it is better to use cut-off
regularization. In a recent paper \cite{Al}, a cut-off $r_c$ in the 
radial coordinate is introduced to regularize the action, which
characterizes
the cusp divergences in a simple way.
In this section, we will compare our numerical results with the cut-off
formula and see that our results numerically agree 
with the cut-off regularized formula of the $n$-point amplitudes.

\subsection{The worldsheet cut-off regularization}
Before going to the radial cut-off formula, we will consider a
simple worldsheet cut-off regularization, where we restrict the 
integration region of $(y_1,y_2)$ in the action (\ref{eq:ngaction}) to 
$[-1+\delta,1-\delta]\times
[-1+\delta,1-\delta]$ for small $\delta$.
Substituting the 4-point solution (\ref{eq:4pt}) into the Lagrangian
$L$, we get
\begin{equation}
 L= {1\over (1-y_1^2)(1-y_2^2)}.
\end{equation}
Here we omit the overall factor $\frac{R^2}{2\pi}$ in the action.
Then the regularized action $S[\delta]$ is defined by
\begin{eqnarray}
S[\delta]&=& \int_{-1+\delta}^{1-\delta}dy_1 dy_2
{1\over (1-y_1^2)(1-y_2^2)}.
\end{eqnarray}
This integral is evaluated easily and we obtain
\begin{eqnarray}
 S[\delta]=\left( \log\frac{2-\delta}{\delta}\right)^2.
\label{eq:wscut1}
\end{eqnarray}
The infrared singularity around the cusp appears as the $(\log \delta)^2$
divergence, which corresponds to the double pole $\frac{1}{\epsilon^2}$
in the dimensional regularization scheme.

\begin{figure}[bthp]
\begin{center}
\resizebox{100mm}{!}{\includegraphics{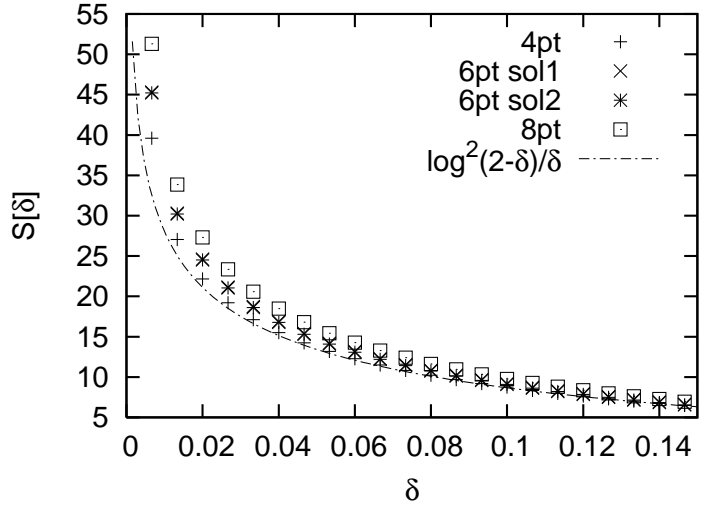}}
\end{center}
\caption{
$\delta-S[\delta]$ graphs at $M=300$.
}\label{fig:dSS}
\end{figure}

We have seen that $S[\delta]$ for cut and glue solutions of 6 and 8-points 
amplitudes is the same as the 4-point amplitude\cite{AsDoItNa}. 
In Fig.~\ref{fig:dSS}, we plot the $\delta-S[\delta]$ graph for 4, 6,
and 8-point 
functions.
{}From the graphs we see that there exist differences among the
regularized actions, especially near the boundaries ({\it i.e.} small 
$\delta$).
This fact indicates that our numerical 6 and 8-point solutions
have different divergent properties from cut and glue solutions.
We see differences between numerical and exact results
for the 4-point amplitude at small $\delta$ region.
These are due to the fact that the numerical results are quite sensitive
with the errors which come from the discretization of $r$ near the boundary.
However, the discrepancy at a given $\delta$ is suppressed
when the lattice size increases, as we will see later.

Since we do not know exact functional form of new numerical solutions,
the $\delta$-dependence of the numerical $S[\delta]$ 
for higher-point amplitudes is not clear.
Roughly we can identify $\log \delta$ as $\frac{1}{\epsilon}$.
It would not be a bad 
approximation to use a 
 fitting function:
\begin{equation}
a (\log \delta)^2+b (\log \delta)+c,
\label{eq:fit1}
\end{equation}
for $S[\delta]$.
Table~\ref{table:ws} is a list of the value of $a$, $b$ and $c$ for each
data fitted in the region $0.04<\delta<1.14$.
The leading singularities are different for 4, 6 and 8 point amplitudes.
But for solution 1 and solution 2 of 6-point functions they agree:
$a_{\rm 4pt}<a_{\rm 6pt sol1}\sim a_{\rm 6pt sol2}< a_{\rm 8pt}$.
This is consistent with the expected infrared leading singularity
behaviour $const. \times n (\log \delta)^2$ for the $n$-point amplitude
\cite{BeDiSm}.

\begin{table}[bthp]
\begin{center}
\begin{tabular}{|c|c|c|c|}
\hline 
& $a$ & $b$& $c$\\
\hline
4-point & 1.12352 & -1.16767& 0.101811 \\
\hline
6-point sol.~1 & 1.40647&-0.648713& 0.214939\\
\hline
6-point sol.~2 & 1.41405 &-0.58385& 0.241626\\
\hline
8-point & 1.67364& -0.26576& 0.280678\\
\hline
\end{tabular}
\end{center}
\caption{
Fitting parameters of $M=300$ data using (\ref{eq:fit1}).
}\label{table:ws}
\end{table}

\begin{figure}[bthp]
\begin{center}
\resizebox{100mm}{!}{\includegraphics{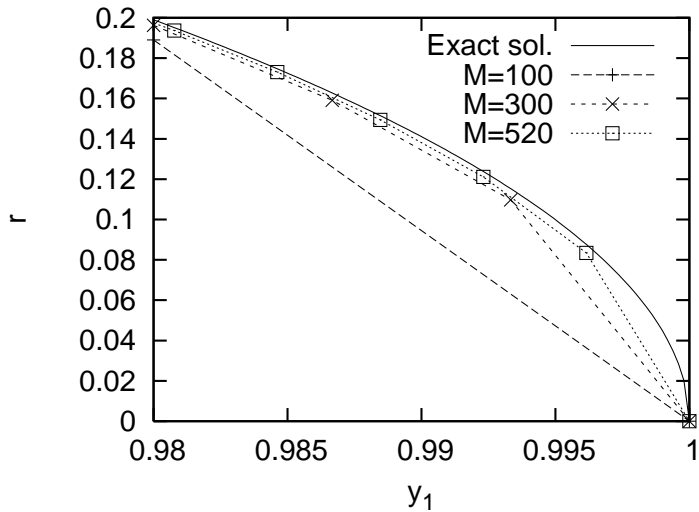}}
\end{center}
\caption{
Near boundary behavior of $r$ on $y_2=0$ plane for several $M$'s
(4-point solution).
}\label{fig:4pt_r_y2eq0}
\end{figure}

\begin{figure}[bthp]
\begin{center}
\resizebox{100mm}{!}{\includegraphics{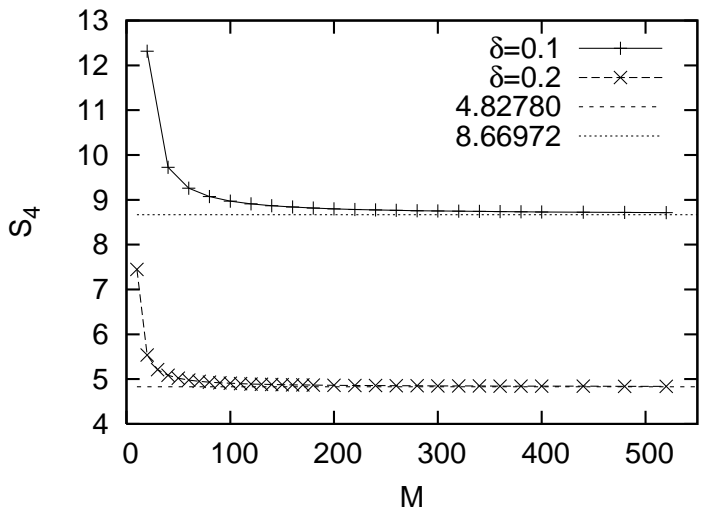}}
\end{center}
\caption{
Regularized area at $\delta=0.1$ and $\delta=0.2$ for several $M$'s
(4-point solution).
}\label{fig:4pt_MSS}
\end{figure}

It is a hard problem to estimate rigorously numerical errors in our 
approximation.
But in the case of the 4-point amplitude,
for which the exact solution has been obtained,
we can see several positive features
which support the validity of our numerical approach.
Fig.~\ref{fig:4pt_r_y2eq0} shows the behaviour of
the numerical solution for $r$ on the $y_2=0$ plane near the boundary.
We can see that the numerical solution approaches 
the exact one as $M$ increases.
Although we omit further details,
it can be also shown that the solution is not affected by
smooth and small modification of the initial condition.
In addition, the area of the surface approaches to
the exact result as $M$ becomes larger and larger.
In Table~\ref{table:4pt_MSS} and Fig.~\ref{fig:4pt_MSS},
we exhibit the behaviours of the cut-off regularized area for 
$\delta=0.1$ and $\delta=0.2$,
which shows good convergence to the analytical results,
with the error being about 0.2\% for the $\delta=0.2$, $M=520$ case.

\begin{table}[bthp]
\begin{center}
\begin{tabular}{|c|c|c|c|c|c|c|c|c|}
\cline{1-3}\cline{5-7}
$M$ & $S_4$& $(S_4-S_e)/S_e$ & 
\hspace*{1em}&
$M$ & $S_4$& $(S_4-S_e)/S_e$ \\
\cline{1-3}\cline{5-7}
100 & 8.97215 & 0.03488390 && 100 & 4.90702 &  0.01641050 \\
\cline{1-3}\cline{5-7}
200 & 8.80018 & 0.01504770 && 200 & 4.86233 &  0.00715303 \\
\cline{1-3}\cline{5-7}
300 & 8.75113 & 0.00939031 && 300 & 4.84934 &  0.00446242 \\
\cline{1-3}\cline{5-7}
520 & 8.71324 & 0.00501923 && 520 & 4.83927 &  0.00237603 \\
\cline{1-3}\cline{5-7}
\multicolumn{3}{c}{(a) $\delta=0.1$}& \multicolumn{1}{c}{} &
\multicolumn{3}{c}{(b) $\delta=0.2$}
\end{tabular}
\caption{
Convergent behavior of cut-off regularized area for the 4-point
 solution.
$S_4$ and $S_e$ are numerical and exact results, respectively.
}\label{table:4pt_MSS}
\end{center}
\end{table}

This artificial worldsheet cut-off is not invariant under conformal
transformation of the worldsheet. 
This regularization  heavily depends on the static gauge.
In order to regularize the action in a conformal invariant way, it is
better to introduce  the radial cut-off as in \cite{Al}.

\subsection{The radial cut-off regularization}
In the radial cut-off regularization scheme we introduce a cut-off $r_c$ in the
radial direction\cite{Al}. 
The regularized area is 
surrounded by the cut-off curve $C: r_c=r(y_1,y_2)$.
For example,
the curve of the 4-point amplitude with $s=t$ is
\begin{equation}
 r_c^2=(1-y_1^2)(1-y_2^2).
\label{eq:cut-off1}
\end{equation}
The regularized action is the area of the region $S$ whose boundary is
$C$:
\begin{equation}
 \tilde{S}[r_c]=\int_S dy_1 dy_2 {1\over (1-y_1^2)(1-y_2^2)}.
\end{equation}
For fixed $y_1$, the integration region of $y_2$ is
$-y^c_2\leq y_2\leq y^c_2$, where $y^c_2(>0)$ is given by
\begin{equation}
 (y^c_2)^2=1-{r_c^2\over 1-y_1^2}.
\end{equation}
After the integration over $y_2$, we get
\begin{equation}
 \tilde{S}[r_c]=\int^{\sqrt{1-r_c^2}}_{-\sqrt{1-r_c^2}}
dy_1{1\over 1-y_1^2}
\log\left(
{\sqrt{1-y_1^2}+\sqrt{1-r_c^2-y_1^2}\over 
\sqrt{1-y_1^2}-\sqrt{1-r_c^2-y_1^2}}
\right).
\end{equation}
Since we are interested in the small $r_c$ limit, we expand the
integrand $f(y_1,r_c)$ in $r_c$:
\begin{equation}
 f(y_1,r_c)=-{2\log r_c-\log(4-4 y_1^2)\over 1-y_1^2}
+{r_c^2\over 2(1-y_1^2)}+\cdots .
\end{equation}
The integral over $y_1$ leads to
\begin{eqnarray}
 \tilde{S}[r_c]={1\over2}\left(\log{r_c^2\over 16}\right)^2
-2(\log 2)^2-{\pi^2\over 16}+O(r_c^2).
\label{eq:4ptactrc1}
\end{eqnarray}

\begin{figure}[bthp]
\begin{center}
\begin{tabular}{cc}
\resizebox{60mm}{!}{\includegraphics{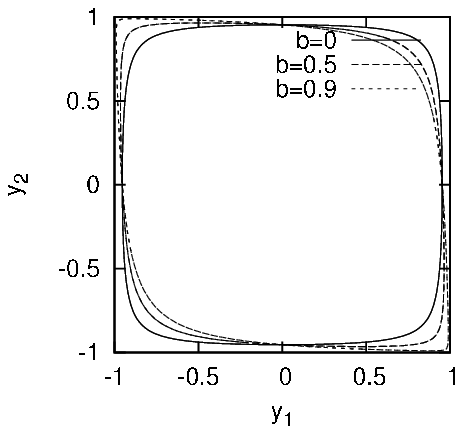}}
&
\hspace{2cm}\resizebox{58mm}{!}{\includegraphics{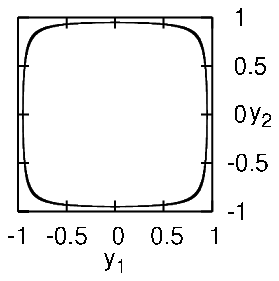}}
\\
\hspace{1.5cm}(a) & \hspace{1cm}(b)
\end{tabular}
\end{center}
\caption{
(a)The cut-off curves with  ${r_c\over a}=0.3$ and $b=0, 0.5, 0.9$
(b) The cut-off curve with ${r_c\over a}=0.3$ 
from the numerical data (4-point solution $M=300$).
}\label{fig:contrcb}
\end{figure}

The general $(s,t)$ solution has
been found in \cite{AlMa}, which can be obtained by scale  and
boost transformation of the $s=t$ solution:
\begin{equation}
 r'={a r\over 1+b y_0},\quad
y'_0={a\sqrt{1+b^2}y_0\over 1+b y_0}, 
\quad
y'_i={a y_i\over 1+b y_0},
\end{equation}
where $a$ is a parameter for the scale transformation and $b$ the boost 
parameter.
The Mandelstam variables $s$ and $t$ are given by
\begin{equation}
(2\pi)^2 s=-{8a^2\over (1-b)^2},\quad (2\pi)^2t=-{8a^2\over (1+b)^2}.
\end{equation}
The cut-off curve (\ref{eq:cut-off1}) 
(see Fig.~\ref{fig:contrcb} (a)) is now replaced by
\begin{equation}
 r_c^2=a^2 (1-y_1^2)(1-y_2^2){1\over (1+b y_1 y_2)^2}.
\end{equation}

We can put $a=1$ by rescaling $r_c\rightarrow r_c a$.
For fixed $y_1$, $y_2$ takes the value in the range
 $y_{2}^{c-} \leq y_2\leq y_2^{c+}$,
where
\begin{eqnarray}
 y_2^{c\pm}&=&{-b r_c^2 y_1\pm \sqrt{(1-y_1^2)(1-y_1^2-r_c^2 +b^2 r_c^2 y_1^2)}
\over 1-y_1^2 +b^2 r_c^2 y_1^2}.
\end{eqnarray}
The integral over $y_2$ yields
\begin{equation}
 \tilde{S}[r_c]=\int_{-\sqrt{{1-r_c^2\over 1-b^2 r^2_c}}}^{\sqrt{{1-r_c^2\over
  1-b^2 r^2_c}}}
dy_1 f(y_1,r_c),
\end{equation}
where
\begin{equation}
 f(y_1,r_c)={1\over 1-y_1^2}{1\over2}\log\left({1+y^{c+}_2\over
                                          1-y^{c+}_2}
{1-y^{c-}_2\over 1+y^{c-}_2}\right).
\end{equation}
Expanding in $r_c$ we get
\begin{eqnarray}
 f(y_1, r_c)=-{1\over 1-y_1^2}\log (r_c^2 {1-b^2 y_1^2\over 4(1-y_1^2)})
+O(r_c^2).
\end{eqnarray}
After the integral over $y_1$, we get
\begin{eqnarray}
 \tilde{S}[r_c]&=& {1\over4}\log^2\left({r_c^2\over -8\pi^2 s}\right)
+ {1\over4}\log^2\left({r_c^2\over -8\pi^2 t}\right)
-{1\over4 }\log^2({s\over t})
-2(\log 2)^2-{\pi^2\over 16}+O(r_c^2).
\nonumber\\
\end{eqnarray}
This formula of the 4-point amplitude was obtained in \cite{Al} using the 
conformal gauge.

The above analysis for the regularized action of the 4-point amplitude
is generalized to examine the infrared singular part of the $n$-point
amplitude, which was done in the dimensional regularization\cite{Bu}.
The infrared singularity of the $n$-point amplitude is 
characterized by the cusp made of two external gluon 
lines with momenta $p_i$ and $p_{i+1}$.
It is convenient to use the light-cone coordinates $y^{\pm}=y^0\pm y^1$
and regard $y_2$ and $r$ as the functions of $y^{\pm}$.
The Nambu-Goto action in this gauge is
\begin{equation}
S={R^2\over 2\pi} \int dy_- dy_+
{1\over 2r^2}
\sqrt{
1-4\partial_- y_2 \partial_+ y_2-4 \partial_- r \partial_+ r
-4 (\partial_- y_2 \partial_+ r -\partial_- r \partial_+ y_2)^2
}.
\end{equation}
The momenta of the external gluon lines in the $(y_-,y_+,y_2)$
coordinates
are parametrized as $2\pi p_i=z_1 (0,1,0)$ and $2\pi p_{i+1}=z_2
(\alpha,1,\sqrt{\alpha})$.
The Mandelstam variable is $(2\pi)^2 s_{i,i+1}=z_1 z_2 (-\alpha)=-z_1
z_2 \alpha$ . 
Then the solution of the equation of motion,
 which approaches to the cusp solution in the limit 
$\alpha\rightarrow 0$, is 
\begin{equation}
 r(y_-,y_+)=\sqrt{2}\sqrt{y_- \left(y_+-{1\over \alpha}y_-\right)},
\quad y_2(y_-,y_+)={1\over \sqrt{\alpha}} y_-.
\end{equation}
Parameterizing the solution as
\begin{equation}
y_-=\alpha z_2 Y_-,\quad
y_+=z_1 Y_++z_2 Y_-,
\end{equation}
we find
\begin{equation}
r(Y_-, Y_+)=\sqrt{2} \sqrt{Y_- Y_+}\sqrt{-(2\pi)^2 s_{i,i+1}},
\end{equation}
and the action near the cusp is
\begin{equation}
S_{i,i+1}={R^2\over 4\pi}\int_0^1 {dY_- dY_+\over 2 Y_- Y_+}.
\end{equation}
The action is divergent at the cusp $Y_{-}=Y_{+}=0$.
Introducing the radial cut-off $r_c$ by
\begin{equation}
r_c=\sqrt{2} \sqrt{-(2\pi)^2 s_{i,i+1}} \sqrt{Y_- Y_+},
\end{equation}
the regularized action rescaled by the factor of ${R^2\over 2\pi}$ becomes
\begin{eqnarray}
 \tilde{S}_{i,i+1}
&=& {1\over2}\int_{{r_c^2\over A^2}}^1 dY_+
\int^1_{{r_c^2 \over A^2 Y_+}} {1\over 2 Y_- Y_+}
\nonumber\\
&=& {1\over8} \left(\log {r_c^2\over
 -2 (2\pi)^2 s_{i,i+1}}\right)^2.
\end{eqnarray}
Here $A=\sqrt{2} \sqrt{-(2\pi)^2 s_{i,i+1}}$.

Then the $n$-point amplitude is expected to have the structure
\begin{equation}
 \tilde{S}_n[r_c]={1\over8}\sum_{i=1}^{n}\left(\log
{r_c^2\over -8\pi^2 s_{i,i+1}}\right)^2
+F_n(p_1,\cdots, p_n)+O(r_c^2),
\label{eq:nptactrc1}
\end{equation}
where
$s_{i,i+1}=-(p_i+p_{i+1})^2$ and $p_{n+1}=p_1$.
 $F_n(p_1,\cdots, p_n)$
is a finite remainder part.
The 4-point function with the momenta (\ref{eq:4ptmom1}) with $s=t$, we
obtain  $(2\pi)^2 s_{1,2}=-8$.
The action is of the form
\begin{equation}
 \tilde{S}_4[r_c]={a_0\over2}\left(\log{r_c^2\over16}\right)^2+c_0,
\label{eq:4ptrc}
\end{equation}
up to $O(r_c^2)$ terms, 
where $a_0=1$ and $c_0$ is a constant. 

For the 6-point function (solution 1 and 2) and the 8-point solution,
the amplitudes in the radial cut-off scheme are
\begin{eqnarray}
 \tilde{S}_{6}^{1}[r_c]
&=&{a_1\over 8}
\left( 4\left(\log {r_c^2\over 8}\right)^2
+\left(\log{r_c^2\over 4}\right)^2
+\left(\log {r_c^2\over 16}\right)^2
\right)+c_1,\label{eq:6pt1rc}\\
 \tilde{S}_{6}^{2}[r_c]
&=& {3\over4}a_2\left(\log{r_c^2\over 8}\right)^2+c_2,
\label{eq:6pt2rc}\\
\tilde{S}_8[r_c]&=&
{1\over2}a_3\left(
\left(\log{r_c^2\over 8}\right)^2+\left(\log{r_c^2\over4}\right)^2
\right)+c_3,
\label{eq:8ptrc}
\end{eqnarray}
up to $O(r_c^2)$ terms,
where $a_1=a_2=a_3=1$ and $c_1,c_2$ and $c_3$ are constants.

\begin{figure}[bthp]
\begin{center}
\resizebox{100mm}{!}{\includegraphics{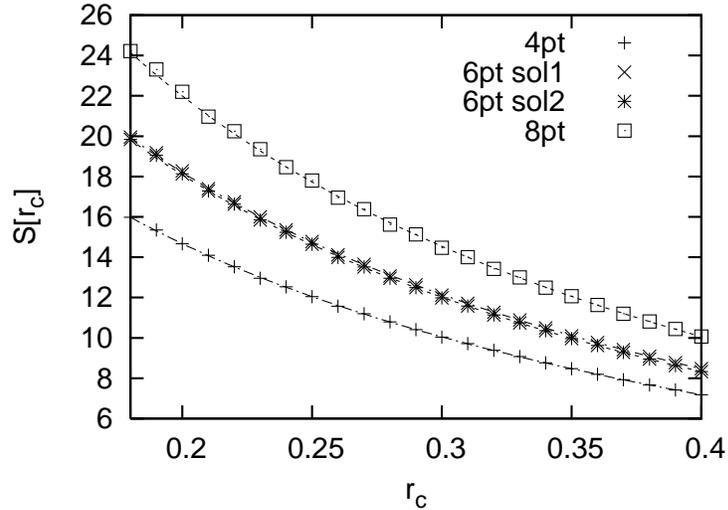}}
\end{center}
\caption{
$r_c$ - $\tilde S[r_c]$ graph at $M=520$.
}\label{fig:rcSS}
\end{figure}

\begin{figure}[bthp]
\begin{center}
\resizebox{100mm}{!}{\includegraphics{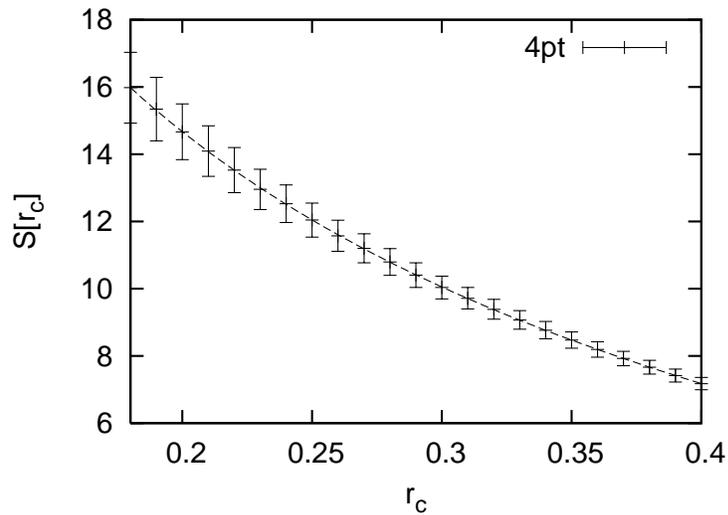}}
\end{center}
\caption{
$r_c$ - $\tilde S[r_c]$ graph at $M=520$ with error-bars.
}\label{fig:4pt_rcSSwe}
\end{figure}

We approximate the area by summing the value of discretized
Lagrangian at the lattice points $(i,j)$ 
at which the value of $r[i,j]$ is larger than the radial cut-off $r_c$.
In Fig.~\ref{fig:rcSS} we show the fitting data of
the $r_c$ regularization.
The numerical results are fitted to the trial functions
with the least-square method
in the range $0.15\le r_c \le 3.0$.
The lower limit of the fitting range cannot be so small 
because of the rapid growth of $r$ near the boundary 
(see Fig.~\ref{fig:4pt_r_y2eq0}).
We reasonably chosen the lower limit as $r_c=0.15$
so that the lattice points around the boundary
which satisfy $r[i,j]>r_c$ are several points inner from the boundary.
The approximation error for the area is roughly estimated as
the difference between the area estimated above
and the one for the nearest outer lattice points.
Fig.~\ref{fig:4pt_rcSSwe} is the result with such errors
for the 4-point solutions.
We fit the results $S[r_c]$ for each solution
with the weight of errors estimated in this way
by using 'fit' command of GNUPLOT.
Fitting these data in the region $0.15\leq r_c\leq 0.3$
by the functions
(\ref{eq:4ptrc}),
(\ref{eq:6pt1rc}), (\ref{eq:6pt2rc}) and (\ref{eq:8ptrc}) respectively,
we find that
\begin{eqnarray}
 a_0&=& 1.01694, \quad c_0=-3.5912,\\
 a_1&=& 1.03715,\quad c_1=-3.68904,\\
 a_2&=& 1.0376,\quad c_2=-3.70123,\\
 a_3&=& 1.04973,\quad c_3=-3.70072.
\end{eqnarray}
We see that our numerical results $a_i$ are in agreement with 
the radial cut-off regularization formula within a few per cent.
The constant $c_0$ of the 4-point amplitude differs from the 
cut-off formula (\ref{eq:4ptactrc1}). 
Therefore we need to take into account for the 
finite $r_c^2$ corrections to the area in order to get the
finite remainder part.
This subject is beyond the scope of the present paper and is left to a
future problem.

\section{Conclusions and Discussion}
In this paper we have investigated the discretized Euler equations 
of the minimal surface in AdS spacetime numerically.
We examined the minimal surface corresponding to the 4, 6 and 8-point 
amplitudes, in which the $n\geq 6$ solutions have the same boundary 
conditions as in
\cite{AsDoItNa}.
These boundary conditions are simple because 
the worldsheet in the static gauge is square, which are easy to write a 
program to solve the system of nonlinear simultaneous equations.
Since the area is divergent around the cusps, we have introduced 
the worldsheet cut-off and radial cut-off regularizations.
In the worldsheet cut-off regularization, the 4-point amplitude agrees
numerically with the exact solution within 0.2\% for the $\delta=0.2$, 
$M=520$ case.
For the radial cut-off regularized amplitudes, we have checked that the
infrared singularity part of the area numerically agrees with the
analytical result near the cusp within 5\% for $520\times 520$ lattice points.

Using conformal transformation of the solution,
we are able to study non-trivial momentum dependence 
of the amplitude including 
not only the infrared singularity but also the finite part.
It is an interesting problem to extract numerically 
the finite remainder part of the gluon amplitudes from the minimal
surface.

\begin{figure}[bthp]
\begin{center}
\begin{tabular}{cc}
\resizebox{75mm}{!}{\includegraphics{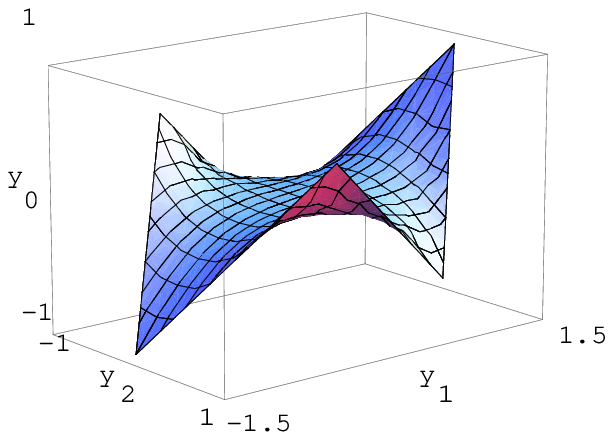}} &
\resizebox{75mm}{!}{\includegraphics{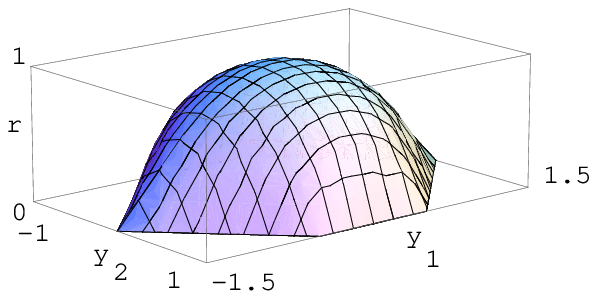}}
\\
(a) & (b)
\end{tabular}
\end{center}
\caption{
Numerical solution of 6-point function with hexagon boundary (116
 lattice points) (a) $y_0(y_1,y_2)$,  (b) $r(y_1,y_2)$.
}\label{fig:hexagon}
\end{figure}

One can also generalize the present square worldsheet to polygons.
Here we will show an preliminary example of the minimal surface 
solution  with 
the hexagonal boundary  
(Fig.~\ref{fig:hexagon}),
whose momenta are 
\begin{eqnarray}
 2\pi p_1&=&(\sqrt{2},1,-1),\quad
2\pi p_2= (-\sqrt{2},-1,-1),\quad
2\pi p_3= (1,-1,0),
\nonumber\\
2\pi p_4&=&(-\sqrt{2},-1,1),\quad
2\pi p_5=(\sqrt{2},1,1),\quad
2\pi p_6=(-1,1,0).
\end{eqnarray}
Here we can see six cusps.
The evaluation of the regularized action with polygon light-like boundary
condition gives
another nontrivial test of
the gluon amplitudes/Wilson loop duality 
at strong coupling.

Another interesting application of this formalism is the gluon
scattering
in non-AdS geometry, which is rather difficult to obtain the
analytic solution of the Euler-Lagrange equations\cite{ItNaIw}. 
A numerical approach will help us to understand their properties
at strong coupling.
A detailed study of these problems  will be discussed elsewhere.

\section*{Acknowledgements}
K. Ito is supported in part by Ministry of Education, Culture, Sports, Science
and Technology of Japan.
K. Iwasaki acknowledges support from the Iwanami Fujukai Foundation.
A part of numerical calculations were performed on Sushiki at YITP,
Kyoto University and the workstation at theory group,  University of
Tokyo, Komaba.

\end{document}